\documentclass[aps,prd,groupedaddress,showpacs]{revtex4}

\usepackage{amsfonts}

\begin{document}


\title{Dual perturbation expansion for a classical $\lambda\phi^4$ field theory}


\author{Marco Frasca}
\email[]{marcofrasca@mclink.it}
\affiliation{Via Erasmo Gattamelata, 3 \\ 00176 Roma (Italy)}


\date{\today}

\begin{abstract}
We prove the existence of a strong coupling expansion for a classical $\lambda\phi^4$
field theory in agreement with the duality principle in perturbation theory put forward in
[M.Frasca, Phys. Rev. A {\bf 58}, 3439 (1998)]. The leading order solution is a
snoidal wave taking the place of plane waves of the free theory. We compute the
first order correction and show that higher order terms do renormalize the leading
order solution. 
\end{abstract}


\maketitle

\section{Introduction}

Quantum field theory is generally formulated to grant a proper weak perturbation
theory starting from a free field. This approach has been proved largely successful as
relevant techniques to cope with infinities and other anomalies following this
approach has been devised \cite{wei1}. Besides, renormalization group methods have
permitted to extract non-perturbative results form these theories \cite{wei2,zj}.

Notwithstanding the long established tradition of a weak perturbation theory, the need
for analysis in the strong coupling limit has become increasingly relevant since
seventies when the discovery of quantum chromodynamics made the situation very difficult
to manage. Bender and colleagues devised an approach for a strong coupling quantum field
theory \cite{bend1,bend2,bend3} but this method proved difficult to apply as all the
relevant quantities depend on a cut-off present in the perturbation series and difficulties
in resumming these series and taking the corresponding limit to zero arose.

Since the pioneering work of Bender and Wu \cite{bw1,bw2} a proper framework to test
strong coupling theories has been the anharmonic oscillator giving also precious
informations about the large order behavior of weak perturbation expansions. Recently,
Kleinert showed as the variational method for generating strong coupling expansions
from a given weak perturbation series, initially devised for the anharmonic oscillator,
produced striking results when applied to a $\lambda\phi^4$ theory to compute critical 
exponents without renormalization group\cite{kle1,kle2,kle3}.

In this paper we follow a different aim. We would like to know explicitly the perturbation
solutions of the field theory as also attempted in the Bender and colleagues works. But
our approach will be somewhat different as it will rely on the recent devised principle
of duality in perturbation theory \cite{fra1,fra2}. This principle is based on the
symmetry observation that the choice of a perturbation term in a differential equation is
arbitrary and one links the series obtained interchanging a term with another. Dual series
having the development parameter inverted are then obtained. So, from a given equation
dual perturbation series can be generated.

We will see that the leading order solution is given by a homogeneous equation and is
represented by a snoidal wave. This leading order solution is renormalized by higher order
terms. The model we consider
will be taken with the mass term always positive. We also limit our considerations to
the two dimensional case even if generalization to higher dimensions is straightforward.
The leading order solution will need a cut-off, that is the space volume should be limited, 
as also happens at the free field theory. Anyhow, classical solutions of the $\lambda\phi^4$
theory display a lot of interesting features otherwise not obtainable and this should also
grant the proper working of a perturbation method whose proof of existence is the main aim of this work.

Quantum field theory in this same limit, $\lambda\rightarrow\infty$, also exists and we
proved this in ref.\cite{fraprd}. We give here a brief description of this theory to prove the fully
consistency and generality 
of our approach. The most important point to be emphasized is that this method yields
the behavior of a quantum field theory in the infrared regime with the relative
mass spectrum.

The paper is structured in the following way. In sec.\ref{sec2} we present the duality
principle in perturbation theory. In sec.\ref{sec3} the dual series is computed for
a $\lambda\phi^4$ field theory showing how the leading order solution gets renormalized
by higher order terms and the first order correction is computed.
In sec.\ref{sec4}, we describe the quantum field theory in the strong coupling limit
as obtained in \cite{fraprd}. 
Finally, in sec.\ref{sec5} conclusions are given.

\section{Duality in perturbation theory and adiabatic approximation\label{sec2}}

In this section we will show how a duality principle holds in perturbation theory
showing how to derive a strong coupling expansion with the leading order ruled by
an adiabatic dynamics in order to study the evolution of a physical system. This
section is similar to the one in \cite{fra3} but we need it in order to make this
paper self-contained.

We consider the following perturbation problem
\begin{equation}
\label{eq:eq1}
    \partial_t u = L(u) + \lambda V(u)
\end{equation}
being $\lambda$ an arbitrary ordering parameter: As is well known
an expansion parameter is obtained by the computation of the series itself. The standard 
approach assume the limit $\lambda\rightarrow 0$ and putting
\begin{equation}
    u = u_0 + \lambda u_1 +\ldots
\end{equation}
one gets the equations for the series
\begin{eqnarray}
    \partial_t u_0 &=& L(u_0) \\ \nonumber 
    \partial_t u_1 &=& L'(u_0)u_1 + V(u_0) \\ \nonumber 
    &\vdots&
\end{eqnarray}
where a derivative with respect to the ordering parameter is indicated by a prime. We recognize here a conventional
small perturbation theory as it should be. But the ordering parameter is just a conventional matter and so one
may ask what does it mean to consider $L(u)$ as a perturbation instead with respect to the same parameter.
Indeed one formally could write the set of equations
\begin{eqnarray}
\label{eq:set}
    \partial_t v_0 &=& V(v_0) \\ \nonumber 
    \partial_t v_1 &=& V'(v_0)v_1 + L(v_0) \\ \nonumber 
    &\vdots&
\end{eqnarray}
where we have interchanged $L(u)$ and $V(u)$ and renamed the solution as $v$. The question to be answered is
what is the expansion parameter now and what derivative the prime means. To answer this question we rescale the
time variable as $\tau = \lambda t$ into eq.(\ref{eq:eq1}) obtaining the equation
\begin{equation}
\label{eq:eq2}
    \lambda\partial_{\tau} u = L(u) + \lambda V(u)
\end{equation}
and let us introduce the small parameter $\epsilon=\frac{1}{\lambda}$. It easy to see that applying again the
small perturbation theory to the parameter $\epsilon\rightarrow 0$ we get the set of equations (\ref{eq:set}) 
but now the time is scaled as $t/\epsilon$, that is, at the leading order the development parameter of the
series will enter into the scale of the time evolution producing a proper slowing down ruled by the equation
\begin{equation}
\label{eq:lead}
    \epsilon\partial_t v_0 = V(v_0)
\end{equation}
that we can recognize as an equation for adiabatic evolution that in the proper limit $\epsilon\rightarrow 0$
will give the static solution $V(u_0)=0$. We never assume this latter solution but rather we will study
the evolution of eq.(\ref{eq:lead}). Finally, the proof is complete as we have obtained a dual series
\begin{equation}
    u = v_0 + \frac{1}{\lambda} v_1 +\ldots
\end{equation}
by simply interchanging the terms for doing perturbation theory. This is a strong coupling expansion
holding in the limit $\lambda\rightarrow\infty$ dual to the small perturbation theory $\lambda\rightarrow 0$
we started with and having an adiabatic equation at the leading order. 

It is interesting to note that, for a partial differential equation, 
we can be forced into a homogeneous equation because, generally, if we require
also a scaling into space variables we gain no knowledge at all on the evolution of a
physical system. On the other side, requiring a scaling on the space variables and not on
the time variable will wash away any evolution of the system. So, on most physical systems
a strong perturbation means also a homogeneous solution but this is not a general rule. As
an example one should consider fluid dynamics where two regimes dual each other can be found
depending if it is the Eulerian or the Navier-Stokes term to prevail. 

\section{Dual perturbation series for a classical $\lambda\phi^4$ field theory\label{sec3}}

We consider a field with the following Hamiltonian
\begin{equation}
    H = \int d^{D-1}x\left[\frac{1}{2}\pi^2+\frac{1}{2}(\partial_x\phi)^2+V(\phi)\right]
\end{equation} 
being $D$ the spacetime dimensionality and $V(\phi)=\frac{1}{2}\phi^2+\frac{\lambda}{4}\phi^4$.
For our aim we will assume $\phi$ real but the extension to a higher number of fields will
have no problems. In this way the Hamilton equations become
\begin{eqnarray}
    \partial_t\phi &=& \pi \\ \nonumber
	\partial_t\pi  &=& \partial_x^2\phi -\phi -\lambda\phi^3.
\end{eqnarray}
We can limit our analysis to the $D=2$ case being  straightforward the extension 
to higher dimensions. Small perturbation theory is easily obtained by taking $\lambda$ as
a small parameter giving the following set of equations
\begin{eqnarray}
    \partial_t\phi_0 &=& \pi_0 \\ \nonumber
	\partial_t\phi_1 &=& \pi_1 \\ \nonumber
    \partial_t\phi_2 &=& \pi_2 \\ \nonumber
	                 &\vdots&  \\ \nonumber
	\partial_t\pi_0  &=& \partial_x^2\phi_0 -\phi_0 \\ \nonumber
	\partial_t\pi_1  &=& \partial_x^2\phi_1 -\phi_1 -\phi_0^3 \\ \nonumber
	\partial_t\pi_2  &=& \partial_x^2\phi_2 -\phi_2 -3\phi_0^2\phi_1 \\ \nonumber
	               &\vdots&				 
\end{eqnarray}
where it easily seen that the free theory, $\Box\phi_0 + \phi_0 = 0$, 
is the leading order solution. Our aim is to
derive a dual perturbation series to this one and, at the same time, to derive the dual
leading order solution to the standard plane waves of the free theory.

In order to reach our aim, following the principle of duality in perturbation theory as
described in sec.\ref{sec2}, we put
\begin{eqnarray}
   \tau &=& \sqrt{\lambda}t \\ \nonumber
   \pi &=& \sqrt{\lambda}\left(\pi_0 + \frac{1}{\lambda}\pi_1 + \frac{1}{\lambda^2}\pi_2 + \ldots\right) \\ \nonumber
   \phi &=& \phi_0 + \frac{1}{\lambda}\phi_1 + \frac{1}{\lambda^2}\phi_2 + \ldots.
\end{eqnarray}
This choice is similar to the one applied in general relativity \cite{fra3}. It is
straightforward to obtain the following not trivial set of equations
\begin{eqnarray}
    \partial_{\tau}\phi_0 &=& \pi_0 \\ \nonumber
	\partial_{\tau}\phi_1 &=& \pi_1 \\ \nonumber
    \partial_{\tau}\phi_2 &=& \pi_2 \\ \nonumber
	                 &\vdots&  \\ \nonumber
	\partial_{\tau}\pi_0 &=& -\phi_0^3 \\ \nonumber
	\partial_{\tau}\pi_1 &=& \partial_x^2\phi_0-\phi_0-3\phi_0^2\phi_1 \\ \nonumber
	\partial_{\tau}\pi_2 &=& \partial_x^2\phi_1-\phi_1-3\phi_0\phi_1^2-3\phi_0^2\phi_2 \\ \nonumber
	                 &\vdots&
\end{eqnarray}
whose solution proves the existence of a dual perturbation series for the classical $\lambda\phi^4$ theory.
At a first glance we notice that the leading order is ruled by a homogeneous equation in space
as already pointed out it could happen in sec.\ref{sec2}. Secondly, we notice that the perturbation
is now the operator $\partial_x^2-1$.  Finally, we have a different time scale that makes 
possible to apply the adiabatic approximation that for this set of equations reduce just to a WKB in time 
due to the fact that we are considering the dual limit $\lambda\rightarrow\infty$ \cite{fra1}. These points
clarify the differences between our approach and the one devised by Bender and colleagues \cite{bend1,bend2,bend3}.

Now we can build the dual leading order solution to the free theory. The leading order is ruled by
the homogeneous equation
\begin{equation}
    \partial_{\tau}^2\phi_0 + \phi_0^3 = 0
\end{equation}
that has the solution
\begin{equation}
    \phi_0 = A{\rm sn}\left[\pm\frac{A}{\sqrt{2}}(\tau+B),i\right]
\end{equation}
being $\rm sn$ the snoidal Jacobi function with modulus $k^2=-1$, and
$A$ and $B$ two integration constants that can be taken dependent on the space
coordinates. Imposing Lorentz invariance we can rewrite the above solution in the
final form
\begin{equation}
    \phi_0 = A{\rm sn}\left[\pm\frac{A}{\sqrt{2}}(\tau+kx),i\right]
\end{equation}
representing a wave with a different period in space and time. In order to normalize
this solution we consider the energy functional given by
\begin{equation}
    E = \int dx\left[\frac{1}{2}(\partial_t\phi)^2+\frac{1}{2}(\partial_x\phi)^2+V(\phi)\right]
\end{equation} 
that at this order would be given by
\begin{equation}
    E_0 = \lambda\int dx\left[\frac{1}{2}(\partial_{\tau}\phi_0)^2+\phi_0^3\right]
\end{equation}
that is unbounded. This means that, as also happens to the free theory, we have to
normalize on a box. So, we take a line with a length $L$ and integrate as
\begin{equation}
    E_0 = \lambda\int^{\frac{L}{2}}_{-\frac{L}{2}}dx
	\left[\frac{1}{2}(\partial_{\tau}\phi_0)^2+\phi_0^3\right]
\end{equation}
that is
\begin{equation}
    E_0=\frac{\lambda}{4}A^4L.
\end{equation}
In turn this means that the values of $k$ are quantized as
\begin{equation}
\label{eq:kn}
    k_n = n\frac{L_0}{L}
\end{equation}
being $L_0=\frac{4\sqrt{2}K(i)}{A}$ an intrinsic length of the system,
$n$ an integer and $K(i)$ the elliptic integral \cite{gr}
\begin{equation}
    K(k)=\int_0^{\frac{\pi}{2}}d\theta\frac{1}{\sqrt{1-k^2\sin^2\theta}}=
	F\left(\frac{\pi}{2},k\right)
\end{equation}
computed for $k^2=-1$. In the same way we can compute the period in time as
\begin{equation}
\label{eq:T}
    T_0=\frac{4\sqrt{2}K(i)}{A}=L_0
\end{equation}
that, as should be expected on the basis of the slowing down in time in a
strongly perturbed system, means $T=T_0/\sqrt{\lambda}$. So, we can rewrite our solution as
\begin{equation}
    \phi_0 = A{\rm sn}\left[\pm 4K(i)\left(\frac{\tau}{T_0}+n\frac{x}{L}\right),i\right]
\end{equation} 
representing the dual solutions to the plane waves of the free theory for the weak coupling
perturbation theory.

The equation to solve at the first order is given by
\begin{equation}
     \partial_{\tau}^2\phi_1 = -k_n^2\phi_0-\phi_0-3\phi_0^2\phi_1.
\end{equation}
To solve this equation we firstly note that the solution can be cast in the form
\begin{equation}
     \phi_1 = \tilde\phi_1-\frac{k_n^2}{2}\phi_0
\end{equation}
leaving us the equation to solve
\begin{equation}
     \partial_{\tau}^2\tilde\phi_1 = -\phi_0-3\phi_0^2\tilde\phi_1.
\end{equation}
Further we notice that due to the adiabatic time scale $\tau$ we do not need to solve exactly
this equation. Rather a WKB solution is enough giving
\begin{equation}
    \tilde\phi_1 = -\frac{1}{2\sqrt{3}}\frac{1}{\sqrt{\phi_0(\tau,x)}}
	\int_0^{\tau}d\tau'\frac{\sqrt{\phi_0(\tau',x)}}
	{\cos\left[\sqrt{3}\int_0^{\tau'}d\tau''\phi_0(\tau'',x)\right]}
	\sin\left[\sqrt{3}\int_0^{\tau}d\tilde\tau\phi_0(\tilde\tau,x)\right].
\end{equation}
The main result we obtain is that the leading order solution gets a correction from the
first order. This situation is the same at the higher orders and we can compute the following
renormalized leading order solution
\begin{equation}
    \phi_{0R}(\tau,x) = \left(1-\frac{k_n^2}{2\lambda}
	-\frac{k_n^4}{8\lambda^2}
	+\frac{5k_n^6}{16\lambda^3}+\ldots\right)\phi_0(\tau,x)
\end{equation}
and the field is renormalized also at the classical level. The renormalization constant
does depend on the cut-off $L$ but in a way that the limit $L\rightarrow\infty$
means the continuous limit for $k_n$. 

We can compute the ground state energy with the renormalized leading order solution. This can be
accomplished by taking into account the integrals
\begin{equation}
\label{eq:i1}
    \int_{-\frac{L}{2}}^{\frac{L}{2}}dx{\rm sn}^2\left[\frac{A}{\sqrt{2}}(\tau+k_nx),i\right]=L
\end{equation}
and
\begin{equation}
\label{eq:i2}
    \int_{-\frac{L}{2}}^{\frac{L}{2}}dx{\rm sn}^4\left[\frac{A}{\sqrt{2}}(\tau+k_nx),i\right]=\frac{L}{3}
\end{equation}
giving rise to the expression for the ground state energy
\begin{equation}
    E_g = Z_E\frac{\lambda}{4}LA^4
\end{equation}
being
\begin{equation}
    Z_E = 1+\frac{2}{A^2\lambda}\left(1-\frac{k_n^2A^2}{3}\right)
	-\frac{2k_n^2}{A^2\lambda^2}\left(1+\frac{k_n^2A^2}{6}\right)+\frac{k_n^6}{\lambda^3}
	+O\left(\frac{1}{\lambda^4}\right)
\end{equation}
where we can see that the leading term of the energy is renormalized by $Z_E=Z_E(\lambda,A,k_n)$.
As it should be expected, 
due to the extensive properties of energy, we
have a direct proportionality with the cut-off $L$. We also note the correct scaling of
energy with $\lambda$. 
Finally we note that the integrals
(\ref{eq:i1}) and (\ref{eq:i2}) could not be as easy to compute in a higher dimensional theory. 

\section{Strongly coupled quantum field theory\label{sec4}}

Quantum field theory in the strong coupling limit, describing the behaviour in the
infrared regime of a $\lambda\phi^4$ theory, has been obtained in \cite{fraprd}. The
starting point is the classical theory obtained sec.\ref{sec3} where is noticed that
the dual perturbation theory can be obtained considering as a perturbation the
term $\partial^2_x-1$. Then, we know that a spatial homogeneous equation rules the
theory at the leading order but we also proved numerically in \cite{fraprd} that in the
limit $\lambda\rightarrow\infty$ the Green function method does hold. Putting all this
together gives the following generating functional
\begin{equation}
    Z[j]=\exp\left[\frac{i}{2}\int d^Dy_1d^Dy_2\frac{\delta}{\delta j(y_1)}(-\nabla^2+1)\delta^D(y_1-y_2)
    \frac{\delta}{\delta j(y_2)}\right]Z_0[j]
\end{equation}
being
\begin{equation}
    Z_0[j]=\exp\left[\frac{i}{2}\int d^Dx_1d^Dx_2j(x_1)\Delta(x_1-x_2)j(x_2)\right]
\end{equation}
with the Feynman propagator
\begin{equation}
    \Delta(x_2-x_1)=\delta^{D-1}(x_2-x_1)[G(t_2-t_1)+G(t_1-t_2)]
\end{equation} 
being
\begin{equation}
\label{eq:gf}
    G(t)=\theta(t)\left(\frac{2}{\lambda}\right)^{\frac{1}{4}}
	{\rm sn}\left[\left(\frac{\lambda}{2}\right)^{\frac{1}{4}}t,i\right].
\end{equation}
In the frequency domain the Feynman propagator is given by
\begin{equation}
    \Delta(\omega)=\sum_{n=0}^\infty\frac{B_n}{\omega^2-\omega_n^2+i\epsilon}
\end{equation}
being
\begin{equation}
    B_n=(2n+1)\frac{\pi^2}{K^2(i)}\frac{(-1)^{n+1}e^{-(n+\frac{1}{2})\pi}}{1+e^{-(2n+1)\pi}}.
\end{equation}
and the mass spectrum of the theory given by
\begin{equation}
    \omega_n = \left(n+\frac{1}{2}\right)\frac{\pi}{K(i)}\left(\frac{\lambda}{2}\right)^{\frac{1}{4}}
\end{equation}
with a mass gap $\delta_0=\frac{\pi}{2K(i)}\left(\frac{\lambda}{2}\right)^{\frac{1}{4}}$.
A proportionality appears between the mass gap and the inverse of the period $T$ given in eq.(\ref{eq:T}).
The quantum theory just fixes the classical amplitude $A$ and the equation for
mass spectrum resembles the one that can be obtained by a WKB formula. 

Application to Yang-Mills theory has been also given obtaining the relative mass spectrum
and mass gap \cite{fraym}. This result can be obtained by mapping the infrared perturbation
theory for the $\lambda\phi^4$ theory described above to the dual perturbation theory for
a Yang-Mills theory. For SU(N) one has 
$\omega_n=\left(n+\frac{1}{2}\right)\frac{\pi}{K(i)}\left(\frac{g^2N}{2}\right)^{\frac{1}{4}}$
with a mass gap given by
$\delta_0 = \frac{\pi}{2K(i)}\left(\frac{g^2N}{2}\right)^{\frac{1}{4}}$,
being $g$ the Yang-Mills coupling constant, that are adimensional being normalized to a constant that is the 
integration constant of a Yang-Mills theory.

Anyhow we point out that in higher dimensions the most relevant aspect should be
the coupling constant renormalization that can make the method not so straightforward
to apply.

\section{Conclusions\label{sec5}}

We were able to show the existence of a dual perturbation solution to the $D=2$ classical
$\lambda\phi^4$ theory and we proved that the dual leading order solutions are snoidal waves. We
also have given explicitly the first order correction and the renormalization of the
leading order solution producing a properly corrected energy.

Besides, in \cite{fraprd} we proved the existence of a infrared quantum field theory
obtained with the same approach applied to the classical theory. This proves the
fully consistency and the generality of this approach in solving partial differential
equations in any situation.

A couple of considerations are in order. Till now the ability to get explicit solutions
to field theory in the strong coupling regime was much limited against the very fine
ability to get high precision values for the relevant observables. This trend may change
with the extension of the above approach to quantum field theory, as already done in general relativity,
opening the way to face physical problems that today are manageable only by numerical
computation.


\end{document}